# Barrier Electrostatics and Contact Engineering for Ultra-Wide Bandgap AlGaN HFETs

Seungheon Shin[1, a)], Can Cao[1], Jon Pratt[1], Yinxuan Zhu[1], Brianna A. Klein[3], Andrew Armstrong[3], Andrew A. Allerman[3], and Siddharth Rajan[1, 2]

[1]Department of Electrical & Computer Engineering, The Ohio State University, Columbus OH, USA
[2]Department of Material Science & Engineering, The Ohio State University, Columbus OH, USA
[3]Sandia National Laboratories, Albuquerque, New Mexico, USA

**Abstract—** We report ultra-wide bandgap (UWBG) AlGaN heterostructure field-effect transistors (HFETs) exhibiting a high breakdown field (> 5.3 MV/cm) and a low contact resistance (~1.55 Ω·mm), tailored for high-power radiofrequency applications. A split-doped barrier architecture, employing two distinct doping concentrations, is shown to enhance both the breakdown field and contact resistance. This design enables a state-of-the-art combination of maximum drain current (487 mA/mm) and breakdown field, along with a high cutoff frequency of 7.2 GHz. These results demonstrate a viable pathway to push device performance toward the material limits while minimizing contact resistance in UWBG AlGaN HFETs, paving the way for next-generation high-power, high-frequency applications.

[a)] Authors to whom correspondence should be addressed
Electronic mail: shin.928@osu.edu

Ultra-wide bandgap (UWBG) AlGaN has emerged as a promising material for next-generation radiofrequency (RF) and millimeter-wave devices due to its superior transport properties, including high saturation velocity and theoretical breakdown field ($F_{BR}$) exceeding 10 MV/cm. These material properties can maximize the Johnson figure of merit (JFOM), defined by the product of breakdown voltage ($V_{BR}$) and cutoff frequency ($f_T$) for RF devices. The JFOM of UWBG AlGaN (> 22 THz·V) is anticipated to significantly surpass that of InP and GaN [1-13]. To realize the highest JFOM, minimizing contact resistance is essential. Zhu et al. reported the lowest contact resistance ($R_C$) of 0.25 Ω·mm in UWBG AlGaN metal-semiconductor field-effect transistors (MESFETs) employing reverse-graded contact layers [7]. In our previous work, we demonstrated UWBG AlGaN HFETs achieving one of the highest breakdown field values (> 5.3 MV/cm) and high maximum drain current ($I_{MAX}$) but a relatively high $R_C$ of 3.26 Ω·mm [10], while employing the same reverse-graded contact scheme used in Zhu et al.'s study [7]. This highlights that achieving a low contact resistance in UWBG AlGaN HFET structures remains a major challenge due to an additional tunneling barrier at the channel and barrier heterostructure interface.

In conventional AlGaN/GaN HEMTs, where the AlGaN barrier is of relatively low composition, annealed ohmic contacts can be used to contact the channel through the barrier [36-40]. For the higher composition barrier, annealed contacts give non-ohmic properties, or relatively high contact resistance [14, 17, 20, 21]. Therefore, in recent work, contact to the channel is made through a doped barrier layer, a reverse-graded layer, and then a metal contact [7, 10, 12, 13]. While a doped barrier layer is important for enabling current injection, it also degrades breakdown and gate leakage. In this work, we introduce a split-doped barrier HFET design and compare its performance to single-doped barrier HFET in terms of on-state, RF, and breakdown performance. We show that the split-doped barrier structure simultaneously improves contact resistance and breakdown robustness, achieving a state-of-the-art combination of breakdown field and maximum drain current in UWBG AlGaN transistors.

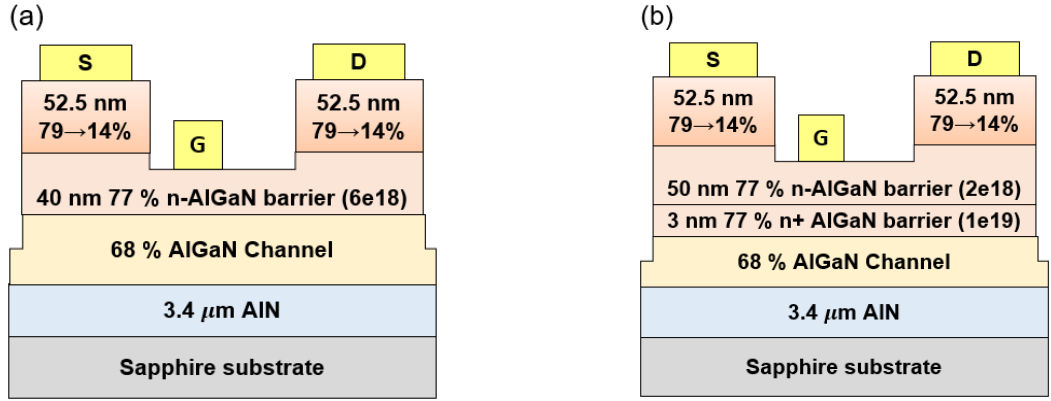

Figure 1. Schematics of epitaxial structures and processed device structures for (a) single-doped barrier HFET, (b) split-doped barrier HFET

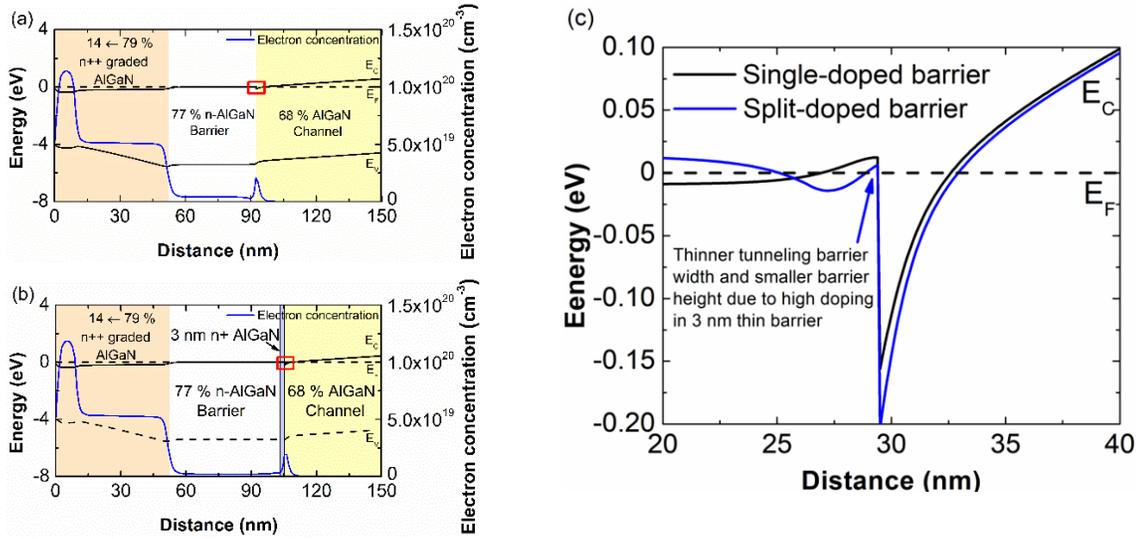

Figure 2. Simulated energy band diagrams for ohmic contact regions at equilibrium condition, cutline from the surface of reverse-graded n++ AlGaN contact layer to a part of the channel layer (a) single-doped barrier HFET structure, (b) split-doped barrier HFET structure, (c) zoomed into red-box in both energy band diagrams and overlayed to describe the effect of split-doped barrier in terms of tunneling barrier reduction

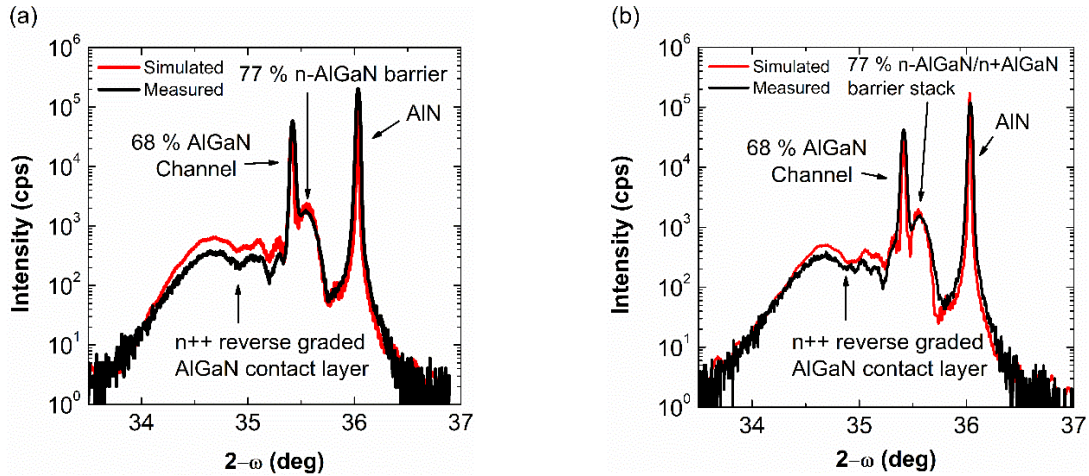

Figure 3. High-resolution x-ray diffraction 2θ-ω scans for (a) single-doped barrier HFET structure [10]. (b) split-doped barrier HFET structure

    To demonstrate the impact of doping design on the device characteristics, two epitaxial layer structures (single-doped (Figure 1(a)) and split-doped barrier HFETs (Figure 1(b))) are reported here. Epitaxial growth was done on pre-deposited AlN/sapphire templates using a TNSC-4000HT metal-organic chemical vapor deposition (MOCVD) reactor. Both structures used $Al_{0.68}Ga_{0.32}N$ channel layers. As shown in the figure, the key difference between these is in the barrier layer between the channel and the reverse-graded contact. The single-doped structure had uniform doping (Si ~ $4 \times 10^{18}$ $cm^{-3}$) [10], while the split-doped employs two doping levels: a lightly doped 50 nm top layer reduces gate leakage and vertical electric field stress across the barrier, enhancing breakdown performance; a highly doped 3

nm bottom layer lowers the conduction band at the barrier/channel interface, improving contact resistance. Simulated energy band diagrams under the ohmic contact (Fig. 2(a)-(c)) illustrate the role of the highly doped 3 nm barrier layer. In the single-doped barrier structure, a thick tunneling barrier and high barrier height at the barrier/channel interface limit electron injection, leading to high contact resistance. In contrast, the split-doped barrier structure exhibits a thinner tunneling barrier and lower barrier height. This is because the highly doped 3 nm layer can increase the tunneling probability and reduce contact resistance by pushing the conduction band down close to the fermi level at the barrier/channel interface (Fig. 2(c)).

High-resolution X-ray diffraction (HR-XRD, Bruker D8 Discover) was used to confirm Al composition and layer thicknesses. Measured spectra matches designed epi-structures, as shown in Fig. 3(a), (b).

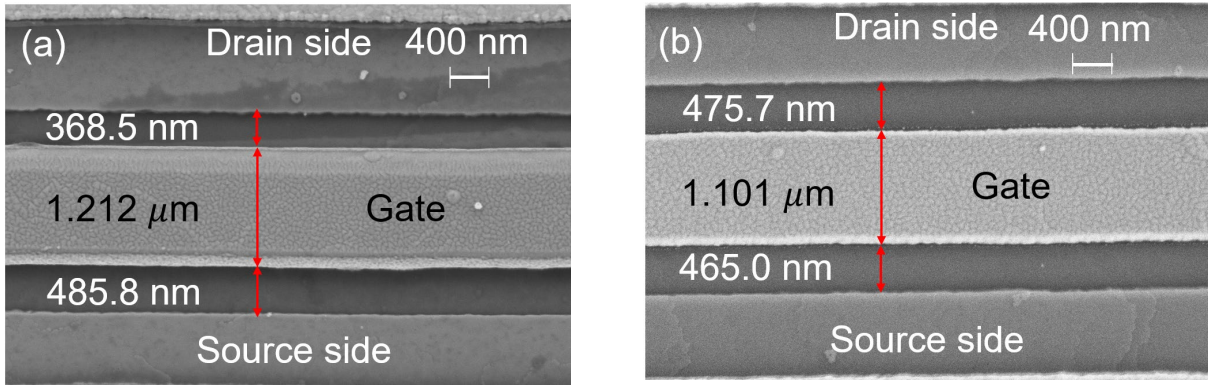

Figure 4. Top view SEM images for processed devices, (a) single-doped barrier HFET, (b) split-doped barrier HFET

Direct-write optical lithography was employed for all patterning steps. To define the access regions, a low-damage ICP-RIE etching process, Ar/BCl$_3$/Cl$_2$ = 5/5/50 sccm, RIE/ICP power = 8/40 W at 5 mTorr, was used, including controlled over-etching into the barrier. The barrier thicknesses after etching were estimated by atomic force microscopy to be about 26 nm and 40.8 nm for the single-doped and split-doped barrier devices, respectively. Non-alloyed ohmic contact (Ti/Al/Ni/Au = 20/120/30/100 nm) was deposited via electron beam evaporation. Mesa isolation was performed using ICP-RIE with an etch depth of approximately 250 nm, followed by gate metal deposition (Ni/Au/Ni = 30/100/20 nm) using E-beam evaporation. The schematics for fabricated devices are shown on Fig. 1. The device dimensions for representative on-state and RF characteristics plots were measured using scanning electron microscopy to be $L_{SD}$ = 2.06 μm, $L_G$ = 1.21 μm, $L_{GD}$ = 0.37 μm for single-doped barrier device, and $L_{SD}$ = 2.04 μm, $L_G$ = 1.1 μm, $L_{GD}$ = 0.48 μm for split-doped barrier device (Fig. 4).

To analyze the ohmic contact properties, transmission line measurements were performed for both epitaxial structures. The single-doped barrier structure showed a relatively high contact resistance of 3.46 Ω·mm and contact resistivity of 2.21 × 10$^{-5}$ Ω·cm$^2$. In contrast, the split-doped barrier structure demonstrated significantly improved values of 1.55 Ω·mm and 5.26 × 10$^{-6}$ Ω·cm$^2$, for contact resistance and contact resistivity, respectively. These results support the expectation based on the energy band diagram, wherein the highly doped 3 nm barrier layer enhances the ohmic contact properties. This indicates the state-of-the-art $R_C$ achieved in UWBG AlGaN HFETs by employing split-doped barrier structure. Hall carrier density and mobility were $7.6 \times 10^{12}$ cm$^{-2}$ and 142 cm$^2$/V·s corresponding to 5.76 kΩ/□ of sheet resistance for single-doped barrier structure, and $1.0 \times 10^{13}$ cm$^{-2}$ and 124 cm$^2$/V·s for split-doped barrier structure indicating 4.79 kΩ/□ of sheet resistance, respectively. The lower mobility in the split-doped barrier structure may be due to the high sheet charge density and remote-ionized scattering from high doping concentration in the thin barrier layer [40, 41].

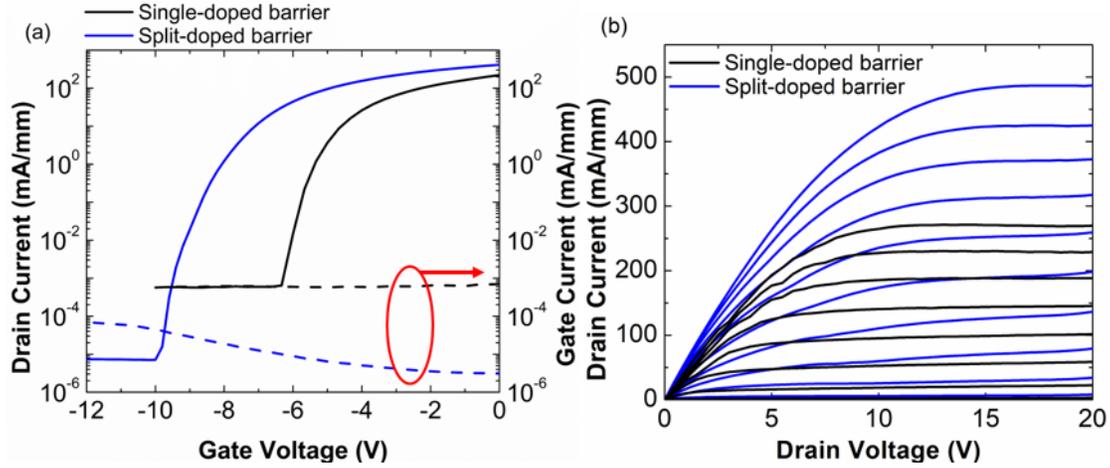

Figure 5. (a) Transfer curves in log scale measured at $V_{DS}$ = 10 V, solid lines: drain current, dashed lines: gate current, (b) output curves with $V_{GS}$ = -8 ~ 2 V, $\Delta V_{GS}$ = -1 V for both epitaxial structures

For DC current-voltage (I−V) characteristics, a Keysight B1500A was utilized. The split-doped barrier device showed a pinch-off voltage ($V_P$) of –10 V due to high total sheet charge density, while $V_P$ for the single-doped barrier device was –6.3 V. A 10× lower gate leakage current ($6 \times 10^{-5}$ mA/mm) was obtained from the split-doped barrier device corresponding to ~ 18× improved $I_{ON}/I_{OFF} > 8 \times 10^6$, supporting that the low-doped thick barrier layer in split-doped barrier structure reduces gate leakage (Fig. 5(a)). By using the split-doped barrier layer, maximum drain current ($I_{MAX}$) was measured to be 487 mA/mm based on improved contact and sheet resistance, while the single-doped barrier devices showed 269 mA/mm (Fig. 5(b)).

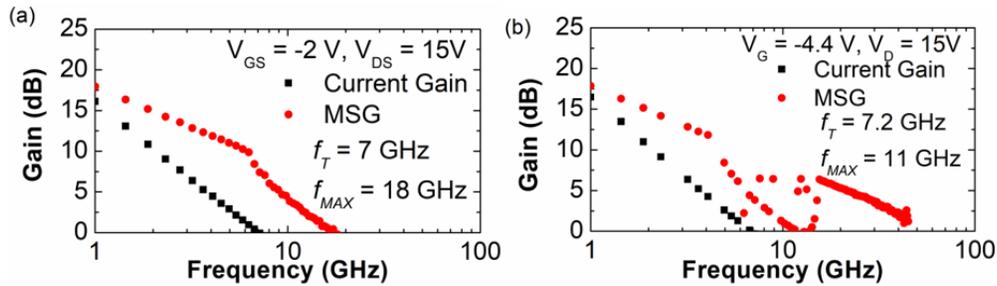

Figure 6. Small-signal measurements for (a) single-doped barrier HFET, (b) split-doped barrier HFET

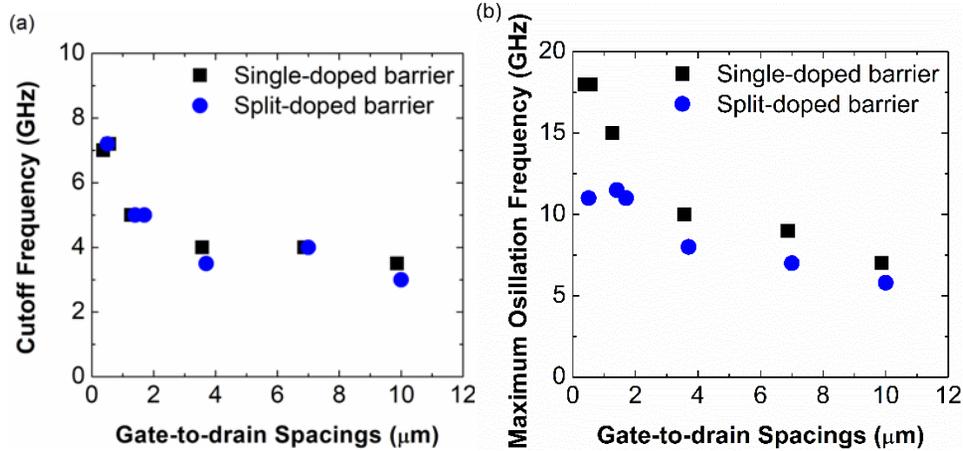

Figure 7. (a) $L_{GD}$-dependent (a) $f_T$ variation, (b) $f_{MAX}$ variation

RF performance was evaluated through on-wafer small-signal measurements using an Agilent 8510C vector network analyzer. All measurements were conducted at the DC bias point that corresponds to the maximum transconductance ($g_m$) on each device. The extracted maximum stable gain (MSG) and short-circuit current gain in dB scale are presented in Fig. 8. The cutoff frequency ($f_T$) and maximum oscillation frequency ($f_{MAX}$) were 7 GHz and 18 GHz for the single-doped barrier device and 7.2 GHz and 11 GHz, respectively for the split-doped barrier device (Fig. 6). The dependence of $f_T$, $f_{MAX}$ on gate-to-drain spacing ($L_{GD}$) is summarized in Fig. 7 to investigate the RF performance trend for longer $L_{GD}$ devices. The limited RF performance in both devices is primarily attributed to contact and sheet resistance. Although the split-doped barrier devices exhibited improved contact resistance, the reduced electron mobility is attributed to limiting $f_T$ performance.

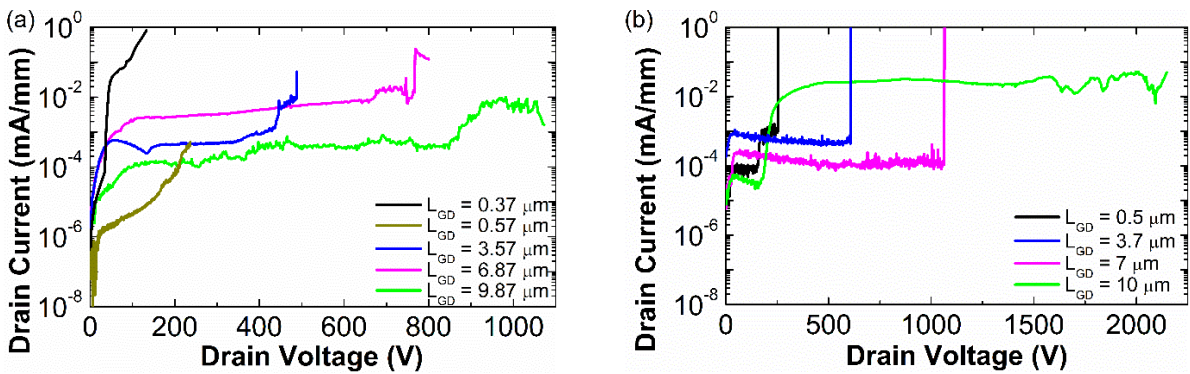

Figure 8. Three-terminal breakdown measurements for (a) single-doped barrier HFETs measured at $V_{GS}$ = -10 V, (b) split-doped barrier HFETs at $V_{GS}$ = -13 V

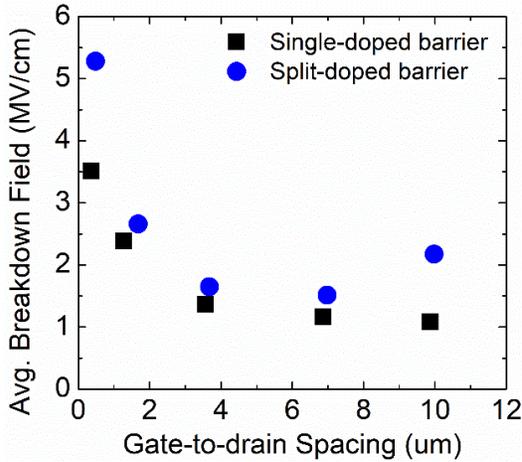

Figure 9. $L_{GD}$-dependent $F_{BR}$ relationship for both devices

For breakdown measurements, three-terminal high voltage tests were conducted via a Keysight B1505A analyzer. The breakdown voltage ($V_{BR}$) in this work was defined as the drain voltage corresponding to a leakage current of 1 mA/mm. The gate bias was set to –10 V for the single-doped barrier devices and –13 V for the split-doped barrier devices, corresponding to $V_{GS} = V_p - 3$. In the split-doped barrier devices, 5.3 MV/cm of high breakdown field ($F_{BR}$) was achieved, which corresponds to the $V_{BR}$ of 253.5 V ($L_{GD} = 0.48$ μm) with on-resistance ($R_{ON}$) of 17.8 Ω·mm (Fig. 8(b)). In addition, $V_{BR}$ of 2147 V and $R_{ON}$ of 78.9 Ω·mm were obtained in 10 μm $L_{GD}$. By contrast, the highest $F_{BR}$ and $V_{BR}$ was 3.5 MV/cm ($L_{GD} = 0.37$ μm) and 1072 V ($L_{GD} = 9.9$ μm) with 81.8 Ω·mm $R_{ON}$ for single-doped barrier devices (Fig. 8(a)). From breakdown measurements, it is suggested that the low-doped thick barrier layer in the split-doped barrier design reduces vertical electric field across the barrier, thereby enhancing breakdown field and robustness. The relationship between $L_{GD}$ and $F_{BR}$ was further explored across different $L_{GD}$ (Fig. 9). Both device types exhibited a decrease in $F_{BR}$ with increasing $L_{GD}$.

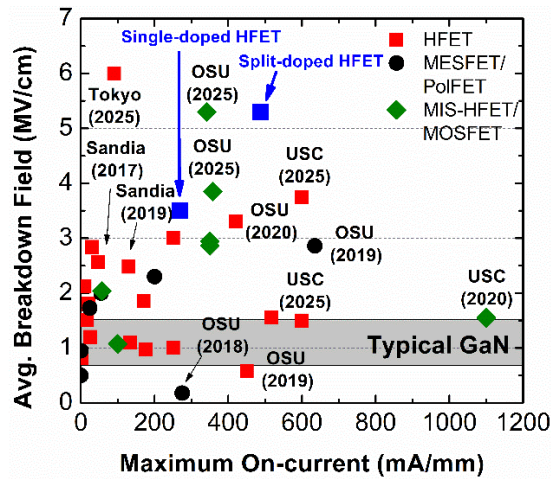

Figure 10. $F_{BR}$ versus $I_{MAX}$ benchmark plot for UWBG AlGaN transistors (channel Al % > 40 %) [8-35]

To summarize the device performance and facilitate comparison with previously reported UWBG AlGaN transistors (channel Al composition > 40%), a benchmark plot of breakdown field versus maximum drain current is presented in Fig. 10 [8-35]. The benchmark plot indicates that split-doped barrier devices achieved a state-of-the-art combination of $F_{BR}$ and $I_{MAX}$, surpassing previously reported values in the same material system.

In conclusion, a split-doped barrier HFET structure was introduced and demonstrated in UWBG AlGaN, exhibiting improved ohmic contact and breakdown performance compared to a single-doped barrier design. By utilizing two doping concentrations in the barrier layers, the lowest reported contact resistance in UWBG AlGaN HFETs of 1.55 Ω·mm was achieved, along with a high drain current of 487 mA/mm, $f_T$ of 7.2 GHz, and a breakdown field exceeding 5.3 MV/cm. The reported results represent a state-of-the-art combination of $F_{BR}$ and $I_{MAX}$ in UWBG AlGaN HFETs, highlighting the potential of this material system to approach its theoretical performance limits.

This work was funded by ARO DEVCOM under Grant No. W911NF2220163 (UWBG RF Center, program manager Dr. Tom Oder). This article has been authored by an employee of National Technology & Engineering Solutions of Sandia, LLC under Contract No. DE-NA0003525 with the U.S. Department of Energy (DOE). The employee owns all right, title and interest in and to the article and is solely responsible for its contents. The United States Government retains and the publisher, by accepting the article for publication, acknowledges that the United States Government retains a non-exclusive, paid-up, irrevocable, world-wide license to publish or reproduce the published form of this article or allow others to do so, for United States Government purposes. The DOE will provide public access to these results of federally sponsored research in accordance with the DOE Public Access Plan https://www.energy.gov/downloads/doe-public-access-plan

**Author Declarations**

**Conflict of Interest**

The authors have no conflicts to disclose.

**Data Availability**

The data that support the findings of this study are available within the article